\documentclass[aps,prl,floatfix,superscriptaddress,showpacs,twocolumn]{revtex4}
\usepackage{amsmath}
\usepackage{graphicx}

\begin{document}
\title{Kerr effect in liquid helium at temperatures below the superfluid transition}
\author{A. O. Sushkov}
\email{alex000@socrates.berkeley.edu} \affiliation{Department of
Physics, University of California at Berkeley, Berkeley,
California 94720-7300} \affiliation{Los Alamos National
Laboratory, Physics Division 23, University of California, M.S.
H803, Los Alamos, New Mexico 87545}
\author{E. Williams}
\email{ewilliam@uclink.berkeley.edu} \affiliation{Department of
Physics, University of California at Berkeley, Berkeley,
California 94720-7300}
\author{V. V. Yashchuk}
\email{yashchuk@socrates.berkeley.edu} \affiliation{Department of
Physics, University of California at Berkeley, Berkeley,
California 94720-7300}
\author{D. Budker}
\email{budker@socrates.berkeley.edu} \affiliation{Department of
Physics, University of California at Berkeley, Berkeley,
California 94720-7300} \affiliation{Nuclear Science Division,
Lawrence Berkeley National Laboratory, Berkeley, California 94720}
\author{S. K. Lamoreaux}
\email{lamore@lanl.gov} \affiliation{Los Alamos National
Laboratory, Physics Division 23, University of California, M.S.
H803, Los Alamos, New Mexico 87545}
\date{\today}
\begin{abstract}
The electro-optical Kerr effect induced by a slowly-varying
applied electric field in liquid helium at temperatures below the
$\lambda$-point is investigated. The Kerr constant of liquid
helium is measured to be $(1.43\pm 0.02^{(stat)} \pm
0.04^{(sys)})\times 10^{-20}$ (cm/V)$^2$ at $T=1.5$ K. Within the
experimental uncertainty, the Kerr constant is independent of
temperature in the range $T=1.5$ K to $2.17$ K, which implies that
the Kerr constant of the superfluid component of liquid helium is
the same as that of normal liquid helium. Our result also
indicates that pair and higher correlations of He atoms in the
liquid phase account for about 23\% of the measured Kerr constant.
Liquid nitrogen was used to test the experimental set-up, the
result for the liquid nitrogen Kerr constant is $(4.38\pm
0.15)\times 10^{-18}\ $(cm/V)$^2$. The knowledge of the Kerr
constant in these media allows the Kerr effect to be used as a
non-contact technique for measuring the magnitude and mapping out
the distribution of electric fields inside these cryogenic
insulants.
\end{abstract}

\maketitle


The electro-optical Kerr effect describes birefringence induced in
an initially isotropic medium by an externally applied electric
field $\vec{E}$ \cite{Ker75i, Ker75ii}. Linearly polarized light
propagating in such a medium experiences a different index of
refraction when its polarization is parallel to $\vec{E}$ compared
to the case when its polarization is perpendicular to $\vec{E}$.
The difference between the two refractive indices is proportional
to $|\vec{E}|^2$:
\begin{equation}
\label {KerrBiref} \Delta n = n_{||}-n_{\perp} = KE^2,
\end{equation}
where $K$ is the Kerr constant of the material. The Kerr constant
is directly related to the third-order non-linear susceptibility.

As far as we know the Kerr effect has not been previously observed
in liquid helium. However the value of $K_{LHe}$ is of importance
both from the fundamental and the applied viewpoints. Kerr effect
measurements probe the nature of the atom distribution and atomic
collisions in the medium under study. In the simplest
approximation of non-interacting spherically symmetric atoms, the
Kerr constant is due to the hyperpolarizability $\gamma$ of an
individual atom, and is proportional to the density of the medium:
$K\propto \rho\gamma$ \cite{Buc55}. This is valid at low
densities, and it is how the measurement of the Kerr effect in
helium gas at room temperature gave $\gamma = (44.2\pm 0.8)\
$atomic units for the helium atom \cite{Boy66, Tam92}. At higher
densities, however, the pair-polarizability anisotropy of the van
der Waals atomic complexes contributes to the Kerr constant as the
second Kerr virial coefficient \cite{Buc68}. The second Kerr
virial coefficient has never been detected for helium, although it
has been measured for the rare gases Ar, Kr, and Xe \cite{Buc68}.
At liquid densities the atoms are so close together that
polarizability anisotropies of two, three, and more atoms all
contribute to the Kerr constant. An interesting question is
whether the superfluidity of liquid helium has any effect on these
many-body interactions and thus the Kerr constant. Furthermore, in
the neighborhood of the helium $\lambda$-transition, anomalous
light scattering due to correlated fluctuations in the order
parameter has been detected, see Ref. \cite{Vin78} for a review.
We pay particular attention to the region around the
$\lambda$-point, looking for any anomalous behavior of the Kerr
constant associated with these fluctuations.

From a more applied standpoint, the Kerr effect can be used as a
non-contact way of measuring the magnitude and the distribution of
electric field in systems where liquid helium is used as the
insulator, such as a superconducting power apparatus \cite{Ger98}.
Specifically, we proposed to use the Kerr effect for monitoring
and diagnostics of the electric field set-up in the new experiment
measuring the parity- and time-reversal - invariance violating
dipole moment of the neutron \cite{LANL}.

\begin{figure}
    \includegraphics[width=\columnwidth]{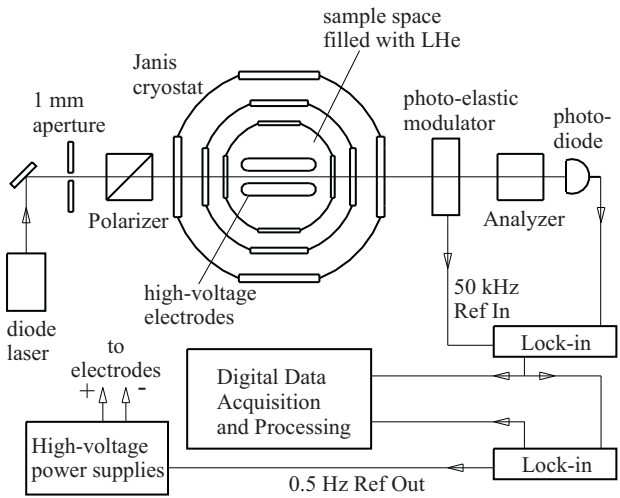}
    \caption{A schematic top view of the experimental apparatus} \label{fig:Apparatus}
\end{figure}
We have observed the Kerr effect in liquid helium using light of
785-nm wavelength, and measured the liquid helium Kerr constant
$K_{LHe}$ in the temperature range 1.5 K to 2.17 K at saturated
vapor pressure. The apparatus used to accomplish this is shown in
Fig. \ref{fig:Apparatus}. We use a Janis model DT SuperVariTemp
pumped helium cryostat with minimum operating temperature of 1.5
K. Optical access is provided via fused quartz windows, the
innermost windows having the diameter of $1/2''$. The central
sample space of the cryostat, holding the liquid helium and the
high-voltage electrodes, is a vertical cylinder, $2''$ in
diameter. At the top of the cryostat, the sample space is
connected to a Stokes Pennwalt rotary vane pump via a
$3/4''$-diameter pumping arm. Also at the top, there is a
connection to an MKS Baratron Type 220B pressure gauge.


In addition, the LakeShore Silicon Diode temperature sensor, model
DT-470-CU-13-1.4L (resolution 0.1 K), is mounted inside the sample
space about 3 cm above the light path. LakeShore model 321
controller, combined with a wound heater mounted next to the
temperature sensor, allows rough temperature control during the
experiment. However we found that recording the helium saturated
vapor pressure with the MKS Baratron allows a much more accurate
measurement (resolution 0.01 K) of the LHe temperature than the
LakeShore temperature sensor.

An electric field is applied across two stainless steel
electrodes, held together by an acrylic fixture, and submerged in
liquid helium. The electrodes are held vertically, in line with
the cryostat windows. Each electrode has the following dimensions:
length = 3.44 cm, height = 3.44 cm, width = 0.88 cm. The edges and
corners of the electrodes are rounded to minimize field
concentration (the residual field-magnitude excess at the edges
over the field in the center of the gap is $\lesssim 5\%$). The
gap between the electrodes is set to $d(293\ $K$) = (0.284\pm
0.001)\ $cm at room temperature. On cooldown the gap shrinks due
to thermal contraction of the acrylic spacers. By illuminating the
gap with a broad light beam and observing the gap image with a CCD
camera, we measured the change in the gap width during a cooldown:
$d(1.5\ $K$) = 0.270\pm 0.003\ $cm.

High voltage is generated by two Glassman power supplies (model
PS/EH20N05.0 and model PS/WG-30P10). Positive voltage is applied
to one of the electrodes, and negative voltage is applied to the
other. Electrical contact between the electrodes and the power
supplies is made through two cryogenic high-voltage cables,
running from the electrodes up through the cryostat sample space,
to feed-throughs at the top of the cryostat.
Because of the $E^2$-dependence of the Kerr effect, it was
desirable to apply the maximum possible voltage to the electrodes,
while avoiding electrical breakdown in the region where the cables
are in contact with helium gas, whose pressure varied from one
atmosphere down to zero. The other challenge in the cable design
was to minimize the heat leak into the liquid helium. Therefore
the conducting portion of the cable was chosen to be a 316L Carbon
steel, $0.051''$-diameter wire. This was insulated with three
layers of Teflon heat shrink tubing, each layer having $0.016''$
wall thickness. At the top of the cryostat each wire is attached
to the feed-through by a brass ball, then a thick layer of curing
silicone glue is spread over any exposed conducting surface to
provide insulation. Tests with the sample space filled with liquid
helium above the $\lambda$-point (under excess pressure to
suppress boiling) show that our high-voltage system allows
application to each electrode of voltages up to 18 kV in
magnitude. With the electrodes separated by 0.27 cm, this
corresponds to the electric field of $130\ $kV/cm. However, when
the liquid helium is at a temperature below the $\lambda$-point,
electrical breakdown at roughly $70\ $kV/cm occurs between the
electrodes. These observations are consistent with the dielectric
strength properties of liquid helium described in the literature
\cite{Ger98}.

We use polarimetry techniques to detect the Kerr effect-induced
birefringence in liquid helium given by Eq. (\ref{KerrBiref}). The
light beam is initially polarized at $45^{\circ}$ to the electric
field. After passing through the liquid helium its polarization
becomes elliptical, with ellipticity
\begin{equation}
\label {KerrEllip} \epsilon = \frac{\pi}{\lambda} K_{LHe} \int E^2
dx = \frac{\pi L}{\lambda} K_{LHe} (V/d)^2.
\end{equation}
Here $\lambda$ is the wavelength of the light, $V$ is the
potential difference between the electrodes, $d$ is the gap, and
the integral over the path of the light beam (taking into account
edge effects) is replaced by the effective electrode length $L$,
which depends on the dimensions of the electrodes as well as their
spacing. A 3-D numerical simulation of the electric field in our
geometry yielded the value of the effective length: $L = (3.20\pm
0.05)\ $cm.

To measure the ellipticity $\epsilon$ we use a modulation
polarimeter (Fig. \ref{fig:Apparatus}). The light source is a
785-nm Hitachi HL7851G laser diode, powered by re-chargeable
batteries, the beam diameter is set to 1 mm by an iris, and the
light power in this beam is 3 mW. The polarimeter consists of a
Wollaston prism polarizer, followed by a Hinds model PEM-FS4
photo-elastic modulator, the second Wollaston prism polarizer
(analyzer), and the photo-diode detector. The photo-elastic
modulator operates at the frequency of 50.2 kHz, the light phase
modulation amplitude is $\pi/4$ and the modulation axis is
parallel to the dark axis of the polarizer. The analyzer is nearly
crossed with the polarizer, it is oriented so that roughly 4\% of
the incident light is transmitted. This de-crossing angle was
chosen to optimize the signal-to-noise ratio. The (non-shot) noise
due to laser-intensity fluctuations scales as the square of the
de-crossing angle, while the ellipticity signal measured by the
polarimeter scales linearly with the de-crossing angle, allowing
such optimization. The sample, whose birefringence properties are
measured, is situated between the first polarizer and the
modulator. With this polarimeter it is possible to simultaneously
measure the ellipticity and the rotation introduced into the light
polarization by the sample. The signal detected by the photodiode
contains harmonics of the modulator frequency (50.2 kHz): the
first harmonic is proportional to the ellipticity, and the second
harmonic is proportional to the rotation. We measure the
ellipticity, using a Stanford Research Systems model SR844 Lock-in
amplifier to detect the amplitude of the first harmonic. To
calibrate, a quarter-wave plate in a precision rotation mount is
inserted into the beam next to the sample, a one degree rotation
of the plate introduces one degree of ellipticity into the light.
The systematic error introduced by this calibration method is 2\%,
due mostly to the uncertainty in the retardation quality of the
quarter-wave plate.

The empty polarimeter has an ellipticity noise of $10^{-7}\
$rad/$\sqrt{\textrm Hz}$. However, when the cryostat is inserted,
the laser beam passes through six  windows introducing an offset
ellipticity, which varies depending on the position of the laser
beam on the windows. The window-induced ellipticity can be $0.2\
$rad or more, when the beam passes near the center of the windows.
We align the laser beam so that it passes close to the edge of the
inside windows, this gives a minimum window-induced ellipticity of
roughly $0.02\ $rad. When the cryostat is operating at liquid
helium temperatures, this offset ellipticity drifts within a range
of about $10^{-4}\ $rad over times of hundreds of seconds. These
drifts are caused by thermal gradients in the window mounts
leading to varying stress applied to the window, and thus to
drifting window birefringence. To separate these drifts from the
Kerr effect - induced ellipticity, the voltage applied to each
electrode is sinusoidally modulated at the frequency of $0.5\ $Hz
(limited by the power supplies), and the Stanford Research Systems
SR830 lock-in (with the integration time set to $10\ $seconds) is
used to pick up the ellipticity signal at the modulation
frequency.

The experimental set-up was tested by measuring the Kerr constant
of liquid nitrogen, which is more than two orders of magnitude
larger than that of liquid helium. At $T=73\ $K our result for the
liquid-nitrogen Kerr constant is: $K_{LN_2} = (4.38\pm 0.15)\times
10^{-18}\ $(cm/V)$^2$. This is in agreement with the published
value of $K_{LN_2}$ \cite{Ima91}. Experimental conditions were
slightly different in that work, but re-scaling the density and
temperature one obtains $K_{LN_2}(T=73$ K$) = 4.32\times 10^{-18}\
$(cm/V)$^2$ from Ref. \cite{Ima91} (no experimental error is
given).

\begin{figure}
    \includegraphics[width=\columnwidth]{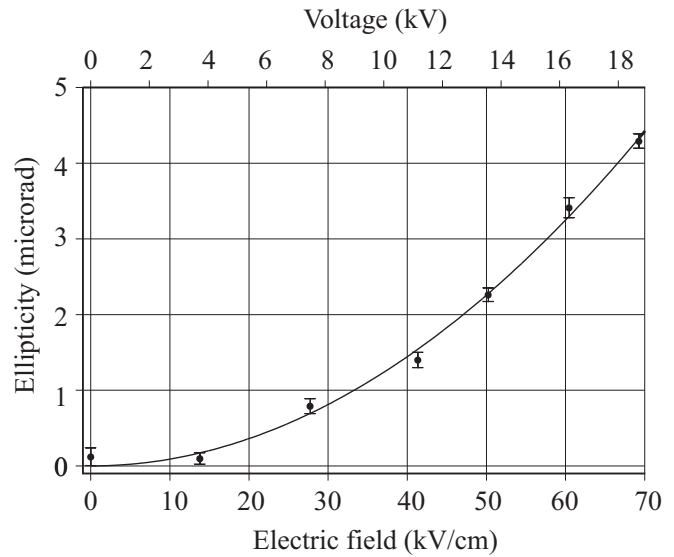}
    \caption{The Kerr effect in liquid helium at $T=1.5\ $K:
    induced ellipticity as a function of the
    electric field between the electrodes.
    Integration time for each point is 1000 s.
    Potential difference between the electrodes
    is shown on the upper scale. Larger electric fields
    were not applied because of breakdown in liquid
    helium.} \label{fig:KerrVsV}
\end{figure}

In liquid helium, as in liquid nitrogen, the Kerr effect was
detected by measuring the light ellipticity $\epsilon$ as a
function of the applied voltage $V$, their relationship is given
by Eq. (\ref{KerrEllip}). The experimental results at $T=1.5\ $K
are shown in Figure \ref{fig:KerrVsV}. A quadratic fit gives the
value of the liquid-helium Kerr constant:
\begin{equation}
\label {KLHe} K_{LHe} = (1.43\pm 0.02^{(stat)} \pm
0.04^{(sys)})\times 10^{-20} \textrm{(cm/V)}^2. \nonumber
\end{equation}
The systematic error is the combination of errors in the electrode
gap, the effective electrode length, and the ellipticity
calibration method. The statistical error is due to
window-ellipticity drifts, and is determined by the integration
time for each point.

It is interesting to compare this number with the Kerr constant
expected from the value of the helium atom hyperpolarizability:
$\gamma = (44.2\pm 0.8)\ $atomic units, deduced from the
measurement of the Kerr effect in the helium gas at room
temperature \cite{Tam92}. The contribution of the
hyperpolarizability of the individual helium atoms to the Kerr
constant of the liquid at $T=1.5$ K is $K^{(0)}_{LHe} = (1.10\pm
0.02)\times 10^{-20}\ $(cm/V)$^2$ \cite{Com01}. There is another,
density-dependent, contribution to $K_{LHe}$, due to the
polarizability anisotropy of van der Waals complexes of two or
more helium atoms in the external electric field. To our
knowledge, this contribution has not been measured for an atomic
liquid. Some theoretical calculations have been performed,
predicting that for liquid argon the contribution due to van der
Waals complexes is a factor of 9 greater than the
hyperpolarizability contribution \cite{Bra69}. The conclusion from
our experimental result for $K_{LHe}$ is that in liquid helium the
contribution due to van der Waals complexes is a factor of 3 less
than the hyperpolarizability contribution. We performed a naive
estimate of the pair-correlation effect, using the pair
polarizability anisotropy of a He$_2$ dimer in the
dipole-induced-dipole approximation, and obtained the right sign
and order of magnitude for the contribution to the liquid-helium
Kerr constant from the pair interactions between the helium atoms.

\begin{figure}
    \includegraphics[width=\columnwidth]{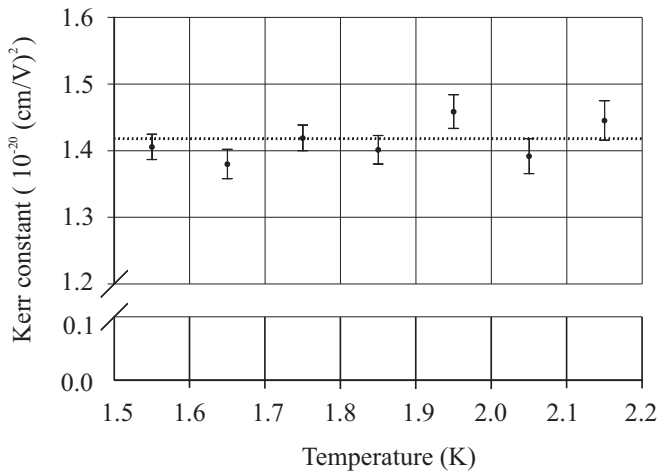}
    \caption{The liquid helium Kerr constant in the temperature range
    $T=1.5\ $K to $T=2.17\ $K. The electric field between the electrodes
    is 69 kV/cm, and the temperature is scanned up or down. Each
    point on the graph is obtained by averaging the Kerr effect - induced
    ellipticity readings over the interval of 0.1 K. The dashed line indicates
    the average over all the points, which agrees with the value of $K_{LHe}$
    given in the text.} \label{fig:KerrVsT}
\end{figure}

The temperature dependence of the liquid-helium Kerr constant was
measured in the range starting at the minimum temperature
achievable in the LakeShore cryostat, $T=1.5\ $K, and up to the
$\lambda$-point, $T=2.17\ $K. As seen from Fig. \ref{fig:KerrVsT},
within the experimental error, the Kerr constant does not depend
on temperature in this range.
This is due, in part, to the fact that the density of liquid
helium changes by only 0.7\% in this temperature range
\cite{Atk59}. However it should also be noted that, in the
two-fluid model, the fraction of the total density that is in the
superfluid component changes from zero at the $\lambda$-point to
about 90\% at 1.5 K. The absence of temperature dependence in Fig.
\ref{fig:KerrVsT} means that, within our experimental accuracy,
the Kerr constant of the superfluid component of liquid helium is
the same as that of the normal component. Furthermore, no
anomalous temperature dependence in the vicinity of the
$\lambda$-point could be seen during the experiment.

No reliable data could be obtained for $K_{LHe}$ above the
$\lambda$-transition. There are several experimental difficulties
encountered when trying to detect the Kerr effect in liquid helium
above the superfluid transition. These are mostly associated with
its low thermal conductivity. Below the $\lambda$-point LHe is a
nearly perfect thermal conductor \cite{Pob96}, therefore there is
no thermal convection or boiling, so the laser beam going through
it is undisturbed. Normal liquid helium, however, is a very poor
thermal conductor ($\kappa \approx 2\times 10^{-4}\
$W/cm$\cdot$K). Even though, to prevent boiling, we kept the
pressure in the cryostat sample space at 1.3 atm, while performing
Kerr-effect measurements above the $\lambda$-point, the laser beam
was still significantly distorted on its passage through LHe,
making the data unreliable. A possible explanation for this is as
follows. Because of the low thermal conductivity and viscosity,
thermal gradients in normal liquid helium create convection
currents and density gradients, which are responsible for the
scattering and distortion of the laser beam. In fact, this
scattering is the basis for optical visualization techniques used
for imaging convective flow patterns in liquid helium
\cite{Woo99}.

In conclusion, we have measured the Kerr constant of liquid helium
to be $(1.43\pm 0.02^{(stat)} \pm 0.04^{(sys)})\times 10^{-20}\
$(cm/V)$^2$ at $T=1.5\ $K, and showed that, within the
experimental uncertainty, it has no temperature dependence in the
range $T=1.5\ $K to $2.17\ $K. The measured value of $K_{LHe}$
indicates that in the liquid phase the polarizabilities of van der
Waals complexes of He atoms account for approximately 23\% of the
experimentally measured Kerr constant. The absence of temperature
dependence suggests that, within the achieved precision,
superfluidity has no effect on the Kerr constant of liquid helium.
Finally, the measurement of $K_{LHe}$ allows the use of Kerr
effect as a non-intrusive technique for measuring the magnitude
and mapping out the distribution of electric fields inside
superfluid liquid helium.


The authors thank S. M. Rochester for help with numerical
calculations, O. P. Sushkov and J. Moore for useful discussions,
and J. C. Davis and R. E. Packard for help with cryogenics. This
research was supported by a UCB-LANL CLE grant, the UC Berkeley
Committee on Research, and by the Los Alamos Directed Research
Grant 2001526DR.

\bibliography{KerrEffectPaper}

\begin{thebibliography}{15}
\bibitem[{Com()}]{Com01}
\bibinfo{note}{We neglect the density-dependence of $\gamma$, since, from the
  results of H. Koch \textit{et al}, J. Chem. Phys. \textbf{111}(22), 10108
  (1999), the correction is on the order of or less than the experimental
  uncertainty.}
\end{thebibliography}

\end{document}